\documentclass[journal,transmag]{./IEEEtran}

\usepackage[printonlyused]{acronym}
\usepackage{amsmath}
\usepackage[pdftex]{graphicx}
\usepackage{enumitem}
\usepackage{cite}
\usepackage{hyphenat}
\usepackage{array}
\usepackage[tight]{subfigure}

\newcommand{\noindento}{\vspace{0.2cm} \noindent}

\begin{document}



\title{Computing and Communications for the Software-Defined Metamaterial Paradigm: A Context Analysis} 
%
%
%
%
%

\author{\IEEEauthorblockN{Sergi Abadal\IEEEauthorrefmark{1},
Christos Liaskos\IEEEauthorrefmark{2},
Ageliki Tsioliaridou\IEEEauthorrefmark{2}, 
Sotiris Ioannidis\IEEEauthorrefmark{2}, 
Andreas Pitsillides\IEEEauthorrefmark{3},\\
Josep Sol\'{e}-Pareta\IEEEauthorrefmark{1},
Eduard Alarc\'{o}n\IEEEauthorrefmark{1}, and
Albert Cabellos-Aparicio\IEEEauthorrefmark{1}}
\IEEEauthorblockA{\IEEEauthorrefmark{1} NaNoNetworking Center in Catalunya (N3Cat),
Universitat Polit\`{e}cnica de Catalunya, Barcelona, Spain}
\IEEEauthorblockA{\IEEEauthorrefmark{2} Institute of Computer Science,
Foundation of Research and Technology-Hellas (FORTH), Heraklion, Greece}
\IEEEauthorblockA{\IEEEauthorrefmark{3} Department of Computer Science, 
University of Cyprus (UCY), Nicosia, Cyprus}%
\thanks{DOI: 10.1109/ACCESS.2017.2693267. 
Corresponding author: S. Abadal (email: abadal@ac.upc.edu).}}

\markboth{IEEE Access}%
{Abadal \MakeLowercase{\textit{et al.}}: Computing and Communications for the SDM Paradigm}

\IEEEtitleabstractindextext{%
\begin{abstract}
Metamaterials are artificial structures which have recently enabled the realization of novel electromagnetic components with engineered and even unnatural functionalities. Existing metamaterials are specifically designed for a single application working under preset conditions (e.g. electromagnetic cloaking for a fixed angle of incidence) and cannot be reused. Software-Defined Metamaterials (SDMs) are a much sought-after paradigm shift, exhibiting electromagnetic properties that can be reconfigured at runtime using a set of software primitives. To enable this new technology, \acp{SDM} require the integration of a network of controllers within the structure of the metamaterial, where each controller interacts locally and communicates globally to obtain the programmed behavior. The design approach for such controllers and the interconnection network, however, remains unclear due to the unique combination of constraints and requirements of the scenario. To bridge this gap, this paper aims to provide a context analysis from the computation and communication perspectives. Then, analogies are drawn between the \ac{SDM} scenario and other applications both at the micro and nano scales, identifying possible candidates for the implementation of the controllers and the intra-\ac{SDM} network. Finally, the main challenges of \acp{SDM} related to computing and communications are outlined. 
\end{abstract}



\begin{IEEEkeywords}
Metamaterials; Software-Defined Metamaterials; Manycores; Approximate Computing; Network-on-Chip; Nanonetworks   
\end{IEEEkeywords}

}

\maketitle
\IEEEdisplaynontitleabstractindextext
\IEEEpeerreviewmaketitle

\acrodef{CMP}{Chip Multiprocessor}
\acrodef{WNSN}{Wireless NanoSensor Network}
\acrodef{WSN}{Wireless Sensor Network}
\acrodef{MAC}{Medium Access Control}
\acrodef{TS-OOK}{Time Spread On-Off Keying}
\acrodef{CSMA}{Carrier Sense Multiple Access}
\acrodef{GWNoC}{Graphene-enabled Wireless Network-on-Chip}
\acrodef{WNoC}{Wireless Network-on-Chip}
\acrodef{NoC}{Network-on-Chip}
\acrodef{TDMA}{Time Division Multiple Access}
\acrodef{FDMA}{Frequency Division Multiple Access}
\acrodef{CDMA}{Code Division Multiple Access}
\acrodef{RF}{Radio-Frequency}
\acrodef{IR}{Impulse Radio}
\acrodef{OOK}{On-Off Keying}
\acrodef{UWB}{Ultra WideBand}
\acrodef{ACK}{ACKnowledgment}
\acrodef{NACK}{Negative ACKnowledgment}
\acrodef{BEB}{Binary Exponential Backoff}
\acrodef{mmWave}{millimeter-Wave}
\acrodef{SDM}{Software-Defined Metamaterial}
\acrodef{SRR}{Split Ring Resonator}
\acrodef{EM}{electromagnetic}
\acrodef{GPGPU}{General-Purpose Graphic Processing Unit}
\acrodef{GALS}{Globally Asynchronous Locally Synchronous}


\section{Introduction}
\label{sec:intro}
\IEEEPARstart{M}{etamaterials} have recently enabled the realization of a wealth of novel \ac{EM} and optical components with engineered functionalities \cite{Engheta2006}. These include \ac{EM} invisibility of objects (cloaking), total radiation absorption, filtering and steering of light and sound, as well as ultra-efficient, miniaturized antennas for sensors and implantable communication devices \cite{Pendry2006, PhysRevApplied.3.037001}. These applications are possible due to the unnatural physical properties of the metamaterials, which stem from their unique structure generally composed of a pattern of conductive material repeated over a 3D volume. If the pattern is replicated over a 2D surface, we obtain a metasurface instead \cite{Glybovski2016, chen2016review}.

Despite its outstanding properties, the adoption of metamaterials and metasurfaces is currently limited due to their non-adaptivity and non-reusability. These properties restrict their applicability to a single functionality per structure (e.g. steering light towards a fixed direction) and to static structures only. Moreover, designing a metamaterial remains a task for specialized researchers, limiting their accessibility from the broad engineering field.

Achieving reconfigurability in metamaterials has been a topic under intense research over the past decade \cite{Oliveri2015}. On the one hand, since the metamaterial properties mostly depend on its conductive pattern, first proposals tried to modulate it using tunable devices or mechanical parts \cite{Hand2007}. On the other hand, more advanced techniques rely on the use of phase-change media, graphene, or liquid crystals \cite{He2015}. The main downturn of these techniques is that the reconfigurability boils down to the tunability of a given static property as there is no actual programmatic control over the functionality. Thus, the accessibility issues are not solved either.

\begin{figure}[!t]
\centering
\includegraphics[width=0.95\columnwidth]{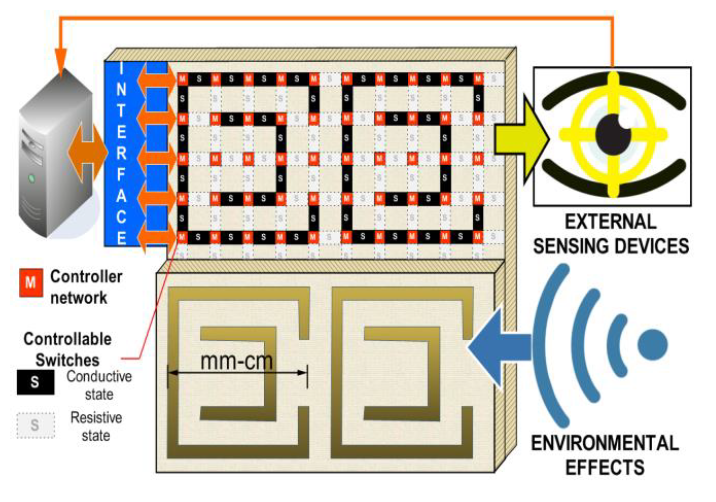}
\caption{Schematic representation of a Software-Defined Metamaterial (SDM). External devices drive a network of controllers, whose local decisions determine the global behavior of the metamaterial.}
\label{fig:SDM}
\end{figure}

Recently, Liaskos {\em et al.} proposed the concept of \acp{SDM}, a hardware platform that can host metamaterial functionalities described in software \cite{Liaskos2015}. The main idea is to integrate a network of miniaturized controllers within the metamaterial structure. The controllers receive programmatic directives and perform simple alterations on the metasurface structure, adjusting its \ac{EM} behavior globally, locally, upon request or depending on the environment. In the specific example of Figure \ref{fig:SDM}, the controllers activate or deactivate their associated switch to determine the metamaterial pattern. The required functionality is described in well-defined, reusable software modules, which are disseminated to the controllers from an external interface also shown in Figure~\ref{fig:SDM}. This has several advantages. First, the \ac{SDM} can host multiple functionalities concurrently and adaptively. Second, the \ac{SDM} can be connected to external devices or even other \acp{SDM} to better adapt to the surroundings or increase the operative range. Last but not least, the \acp{SDM} concept reduces the knowledge required to design a metamaterial for a given purpose.

As mentioned earlier, a network of controllers lies at the heart of an \ac{SDM}. Both the controllers and their interconnections would ideally be simple, ultra-efficient, yet powerful enough to enable real-time adaptivity and support multiple ways of interacting locally, globally, and with external entities. However, this combination of constraints and requirements poses important challenges, thus requiring a careful definition of the computation and communication mechanisms that will drive the operation of \acp{SDM}. 

This position paper aims to provide a context analysis of the \ac{SDM} paradigm from the computing and communication perspectives. We build on the observation that existing approaches may be amenable to this new application if adapted properly. As the main contribution, this work does not aim to deliver a working solution, but rather: 
\begin{itemize}   \itemsep1pt \parskip1pt \parsep1pt
\item To provide a broad analysis of the application context, detailing its particularities regarding the physical implementation, workload characteristics and performance requirements.
\item To present an overview of existing computing and networking approaches that could be amenable to \acp{SDM}.
\item To enumerate the outstanding challenges of this new research area, paving the way for future investigations.
\end{itemize}  

The remainder of this paper is organized as follows. Section~\ref{sec:metamaterials} provides background on the reconfigurable metamaterial paradigm and analyzes its main particularities. Then, Section~\ref{sec:computing} debates the applicability of current computing techniques to the \ac{SDM} scenario. Sections~\ref{sec:contextNoC} and~\ref{sec:contextOther} extend the discussion to the networking domain in general and the \ac{NoC} paradigm in particular. Finally, Section~\ref{sec:challenges} lays out the main computation and communication challenges of \acp{SDM} and Section~\ref{sec:conclusions} concludes the paper.

\section{Software-Defined Metasurfaces} 
\label{sec:metamaterials} 
For simplicity, let us focus on a particular 2D metasurface case shown in Figure \ref{fig:SDM}. In this case, the dimensions of the rectangular \acp{SRR} define the refraction angle of an impinging \ac{EM} wave. Each controller is associated to a switch (or a set of switches) that can be set on conductive or resistive state, therefore shaping the \acp{SRR} used as building blocks. Changes of state in each switch can be prescribed via the metasurface interface either because the user desires to change the refraction angle or because external sensing devices detect changes in the \ac{EM} source. The scale of the controllers and the switches defines the granularity of the formable patterns, eventually determining the number of possible configurations and the frequency at which the metasurface can operate. We refer to the interested reader to \cite{Liaskos2015} for more details.

\noindento {\bfseries General structure of an \ac{SDM}.} The particular example of Figure \ref{fig:SDM} represents one of the different potential approaches that can be used to attain reconfigurability in an \ac{SDM}. Other schemes may involve the use of tunable resistors or capacitors, the value of which determines the behavior of the \ac{SDM} and is dictated by the controller. With the use of graphene, which is inherently tunable, an \acp{SDM} can be created by allowing controllers to change the electrostatic bias applied to the different areas of the graphene sheet. Regardless of its physical characteristics, a generic instance of an \ac{SDM} would have the logical structure shown in Figure \ref{fig:struct}:

\begin{figure}[!t]
\centering
\includegraphics[width=\columnwidth]{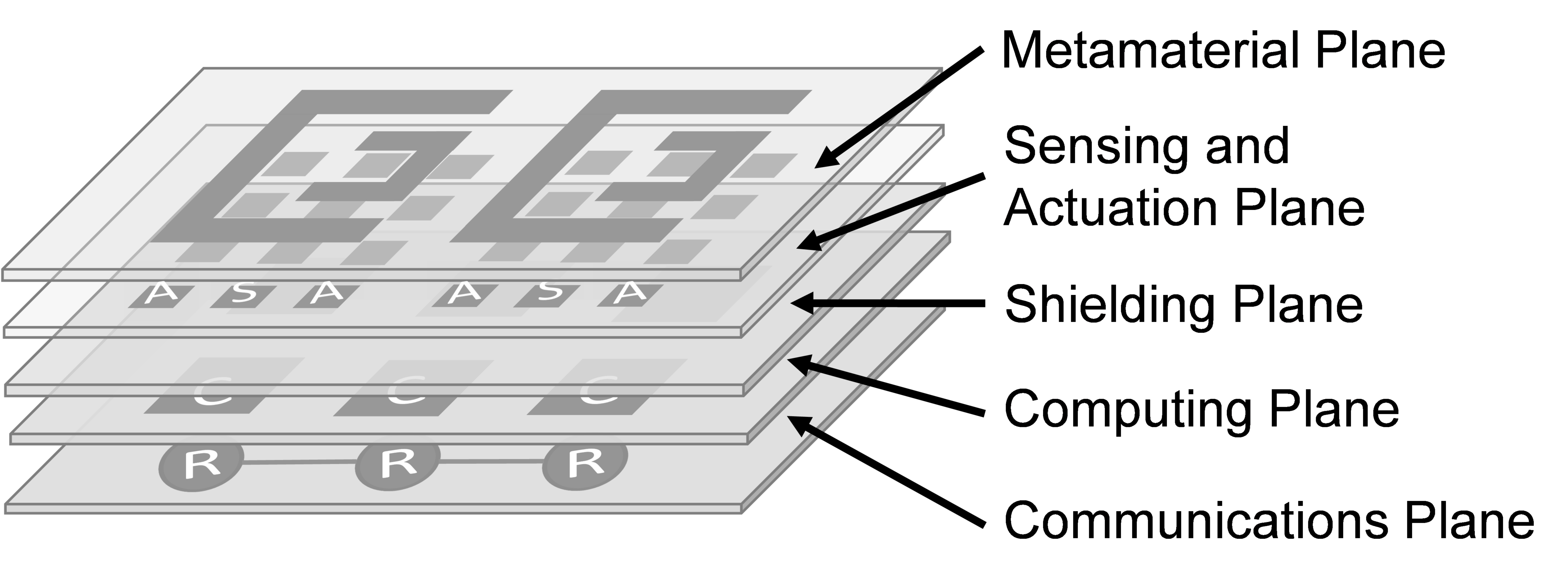}
\caption{Sketch of the logical structure of an \ac{SDM}. It may includes actuators (A), sensors (S), controllers (C) and routers (R).}
\label{fig:struct}
\end{figure}

\begin{itemize}  \itemsep1pt \parskip1pt \parsep1pt
\item {\bf Metamaterial Plane:} which delivers the desired \ac{EM} behavior through a reconfigurable pattern. The metamaterial plane can be implemented, for instance, with CMOS switches as illustrated in Figure \ref{fig:SDM} or materials such as graphene, which can be tuned by simply changing an electrostatic bias \cite{He2015}.
\item {\bf Sensing and Actuation Plane:} which modifies the behavior of the metamaterial plane. Successive \acp{SDM} generations may integrate sensors within the metasurface, so that state changes can be determined internally without the need to reach an external controller, thereby providing a truly autonomous and adaptive operation.
\item {\bf Shielding Plane:} which attempts to decouple the \ac{EM} behavior of the top and bottom planes, aiming to avoid mutual interferences. A simple metallic layer could be used to this end, as metals mainly reflect \ac{EM} waves.
\item {\bf Computing Plane:} which executes external commands from the interface and internal commands from the rest of controllers or sensors to effectively change the \ac{EM} profile of the metamaterial plane. Note that one controller can drive the operation of one or several actuators. Possible design approaches are discussed in Section~\ref{sec:computing}.
\item {\bf Communications Plane:} which coordinates the actions of the computing plane and keeps in touch with external entities via the \ac{SDM} interface. It may be wired or wireless. Possible design approaches are discussed in Sections~\ref{sec:contextNoC} and~\ref{sec:contextOther}. 
\end{itemize}

At this point, it is important to stress that the programmability of \acp{SDM} refers to their \ac{EM} properties only. This differentiates \acp{SDM} from the Claytronics project, which aims to program changes in the physical shape of matter \cite{Claytronics}. In any case, we will later see that advances in that application context can be meaningful to the \ac{SDM} paradigm as they share some basic traits. 

\begin{figure}[!t]
\centering
\includegraphics[width=\columnwidth]{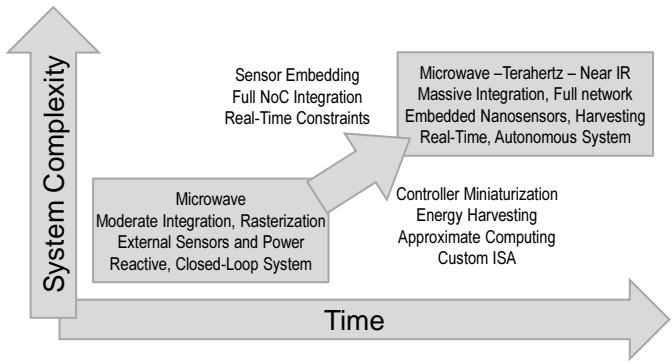}
\caption{Current perspectives and expected evolution of the \ac{SDM} research activities.}
\label{fig:timeline}
\end{figure}

\noindento {\bfseries Current Perspectives and Vision.} The potential of the \ac{SDM} concept is vast given the plethora of potential applications in the microwave range and above. However, their feasibility is currently limited to the development of proof-of-concept devices maintaining a simple architecture and performance. As shown in the left part of Figure  
\ref{fig:timeline}, those initial \acp{SDM} would be limited reactive systems in the microwave range with external sensing and power supply. In the longer term, the \ac{SDM} vision could incorporate new components such as embedded nanosensors, a full integrated network, or an energy harvesting system, and exploit smaller and faster controllers to create devices capable of reacting microwave or terahertz signals in a truly autonomous manner, without having to rely on the constant intervention of an external controller.

\noindento {\bfseries Context Analysis.} In the following, the main characteristics of the \ac{SDM} application are analyzed considering both the current state of things and the full potential of the \ac{SDM} vision. The main insights are summarized in Table \ref{tab:scenario}.

\begin{table*}[!t] 
\centering
\caption{Communications and Computing in the SDM scenario}\label{tab:scenario}
\begin{tabular}{m{4cm}m{6cm}m{6cm}} 
Perspective & Traits & Analogous Techniques \\ \hline\hline
Physical Landscape & Planar, dense, constrained, static, controlled (spatially periodic) & Embedded manycores, GPGPUs, NoCs, WNoCs, Nanonetworks, Energy harvesting \\ \hline
Workload Characteristics & Light, highly correlated (dissemination, reductions, sensing), monolithic nature & Multicore processors, NoCs, WNoCs, Sensor networks \\  \hline
Application Requirements & Depends on granularity, from latency-insensitive to real-time, tolerant to errors & Mission-critical systems, Nanonetworking, Approximate computing \\ \hline\hline
\end{tabular}
\end{table*}

\subsection{Physical Landscape}
\label{eq:physical}
Computing and communications occur within a constrained environment. The lateral dimensions of the metamaterial building blocks are generally $\lambda$/4 or less, where $\lambda$ is the wavelength of the \ac{EM} waves impinging on the metamaterial. This, for the example of Fig.~\ref{fig:SDM}, means that a reasonable target of $f=6$ GHz would require the deployment of an \ac{SRR} every $\sim$1 cm. Assuming that each \ac{SRR} is composed by dozens of switches, controllers would be placed every $\sim$1 mm approximately. Note that such density requirements can be relaxed if concentration is applied, i.e., each controller is shared by a few switches. It is also worth noting that the controllers will operate at a frequency generally much lower than that of the manipulated \ac{EM} waves.

The granularity of actuation scales inversely to frequency, therefore generating a considerably dense and highly integrated network of as-small-as-possible controllers. Due to this density and to minimize heat and potential interferences, both the controllers and the network should have a strict power budget also related to the frequency of the impinging \ac{EM} waves. Link energy figures in \ac{NoC}, currently in the pJ/bit range and below, can serve as a first reference. In future systems where the \ac{SDM} is meant to be autonomous and powered by the same \ac{EM} source than that the controlled by the metamaterial, the energy budget should comply with the limitations of the energy harvester.

The computing and communications devices will be laid out in a planar environment, probably in a chip-like configuration, if we consider the metasurface case; whereas this should not be necessarily the case in the broader sense of the \ac{SDM} paradigm. In both cases, however, the topology of actuators reconfiguring the pattern will be static, controlled, and known beforehand (most likely fairly periodic). As we will see, this offers important optimization opportunities. 

\subsection{Workload Characteristics}
\begin{figure*}[!t]
\centering
\includegraphics[width=0.7\textwidth]{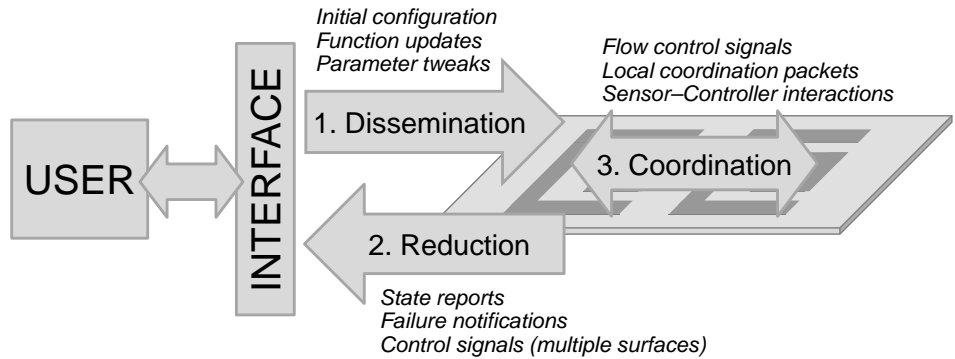}
\caption{Computation and communication flows in \acp{SDM}.}
\label{fig:patterns}
\end{figure*}

Although the \ac{SDM} paradigm opens the door to a large wealth of possibilities at the metamaterial plane, the computing and communication planes only need to perform three distinct actions, summarized in Figure \ref{fig:patterns}:
\begin{enumerate} \itemsep1pt \parskip1pt \parsep1pt
\item Receive and execute external directives. This basically implies the dissemination of data from the interface to all the controllers and the execution of (preferably state-independent) instructions for the initial configuration of the metasurface and the subsequent function updates. After receiving feedback from external sensors or the metasurface itself, the interface may also need to convey messages containing parameter adjustments required to maintain the desired behavior.
\item Process and send internal information to the interface. For debugging or \ac{SDM} interconnectivity purposes, controllers may need to individually or collectively communicate with the interface, therefore generating a \emph{reduction} operation with temporally correlated many-to-one traffic. In the former case, the metasurface will send periodic state reports or sporadic failure notifications. In the latter case, the interface will receive control signals from the different metasurfaces in order to coordinate their joint operation.
\item Coordinate their execution strictly within the \ac{SDM}. To maintain the correct behavior of the \ac{SDM}, integrated sensors may need to communicate with the controllers and drive their execution. These events generate point-to-point or multicast communication with potentially high spatial correlation. Controllers may also need to locally notify errors and perform flow control within the network.
\end{enumerate}

On top of these considerations, it is important to note that the communication and computation intensity will end up depending on the desired spatial and temporal granularity, as well as on the variability of the \ac{EM} waves impinging on the \ac{SDM}. In any case, given the nature of the application and of the energy constraints of the controllers, the load should be moderate. 

Another interesting point is that the \ac{SDM} will be a monolithic system, meaning that designers will have control over the entire architecture, from the physical implementation up to the compilers. This may have little impact on the computing side since multiprocessors are generally monolithic as well. However, it represents a big departure from traditional networks where the nodes, protocols, and applications are developed by different teams. This implies that protocols can be streamlined by entering into the design loop of the whole architecture as in \acp{NoC}.

\subsection{Application Requirements} \label{appReq}
The requirements set by the application mostly depend on the desired spatiotemporal granularity. In the first \ac{SDM} generations, where the main objective is to attain reconfigurability via software, latency requirements are expected to be relaxed, probably between a few milliseconds and a few seconds. In a longer term, where \ac{SDM} applications may demand fast adaptivity, stronger timing requirements on the order of microseconds may be imposed to the controllers and the network. Designs will favor simplicity against performance in the former case, while real-time constraints will suggest the use of mission-critical solutions in the latter case. 

An interesting feature stemming from the fundamentals of the \ac{SDM} application concerns the reliability requirements. Depending on the particular design of the metamaterial pattern, the task of the controller may be, for instance, the choice of a discrete set of voltage levels. The failure of a few controllers or the choice of an incorrect voltage level may not be noticed at the macroscopic level, still obtaining the desired \ac{EM} behavior. This situation can be quantified and used to improve the efficiency of the controllers and the network. 

\section{Applicability of Current Computing Trends}
\label{sec:computing}
The analysis of the \ac{SDM} context has clarified that the computing plane will be massive, composed by a potentially huge amount of tiny controllers deployed within a single monolithic system. As a result, simplicity will most likely drive the development of controllers and lead to custom solutions. Each controller will have to handle commands from external entities or from internal controllers or switches, to compute the new state of its associated switches or actuators. This operation is required to obtain the desired feature (e.g. a given pattern, impedance, bias) in the pathway to obtain the target macroscopic \ac{EM} behavior.  

\begin{figure}[!t]
\centering
\includegraphics[width=\columnwidth]{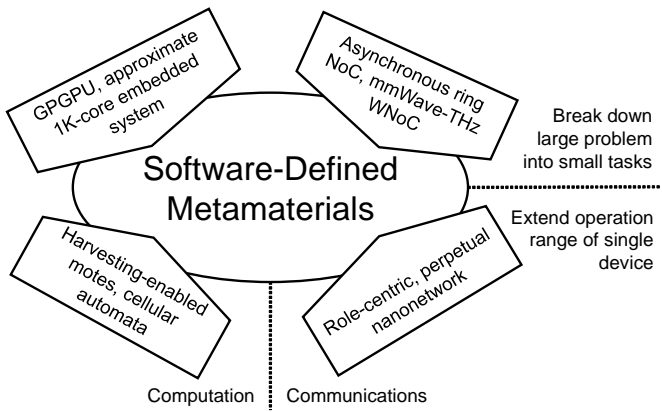}
\caption{The \ac{SDM} scenario seen from the perspective of possible computation and communications solutions.}
\label{fig:SDMconc}
\end{figure}

In the following, we revisit how the \ac{SDM} community can benefit from existing knowledge in the area of computing. The main conclusions are summarized in the left part of Figure \ref{fig:SDMconc}.

\subsection{Massive Manycores} \label{src:manycore}
Taking into consideration their density and \emph{a priori} monolithic nature, the network of controllers within \acp{SDM} can be seen, at large, as a massive manycore processor. Such processors already exist in the research domain, reaching the thousand-core count within a single chip not only in theoretical discussions \cite{Borkar2007}, but also built and demonstrated with CMOS technology \cite{Bohnenstiehl2016}. Strictly speaking, however, an \ac{SDM} does not include a general multiprocessor, but rather an embedded manycore as it can be described as a \emph{computing system with a dedicated function within a larger mechanical or electrical system, often with real-time constraints.} Development of \ac{SDM} controllers can therefore inherit experience of past custom architectures or software with real-time constraints for embedded multicores, especially considering that they are already used in other software-defined paradigms \cite{Choi2009}.

In their work, Liaskos {\em et al.} discuss the suitability of massively parallel computing architectures mostly due to their node density and the fact that all controllers perform a small set of identical functions \cite{Liaskos2015}. \acp{GPGPU} such as CUDA-enabled video cards are mentioned as they can handle thousands of threads, conveniently organized in sets and executing simple operations \cite{Owens2007}. The possible use of \acp{GPGPU}-like computing organizations, at least for proof-of-concept explorations, may be backed up by the vast amount of applied research and knowledge gained through the widespread adoption of these devices in the scientific domain. 

%

\subsection{Towards Infinitesimal Computing}
\label{sec:infinitesimal}
The top-down view of an \ac{SDM} implicitly assumes that a large task is divided into multiple and possibly identical subtasks to reach a common goal. This matches well with the process of reconfiguring the \ac{SDM} via the software interface. More prospectively, if we envision allowing \ac{SDM} to internally sense and adapt to different \ac{EM} conditions, a bottom-up perspective might be more appropriate.

In strict terms, the controllers and the associated integrated sensors (if any) form a sensor and actuator network \cite{Melodia2007}. One controller is not significant by itself as it can only impact on one or a few building blocks of the metamaterial, and therefore needs to be connected to other controllers to obtain a desired macroscopic behavior.

Regarding node density and size limitations of controllers, \acp{SDM} are conceptually close to paradigms such as smart dust \cite{Warneke2001}, Claytronics \cite{Claytronics}, or \ac{WNSN} \cite{Jornet2010}. The potentially infinitesimal motes or nanorobots forming these networks account for tiny computing capabilities and may need energy harvesting modules to operate. Thus, existing knowledge on how to develop and program these systems, e.g. using an event-centric approach, may be highly relevant to the \ac{SDM} community \cite{Levis2005}.

Finally, it is worth noting that the periodic layout and simplicity requirements of \acp{SDM} allows us to draw a very strong analogy to the cellular automata approach \cite{Sarkar2000}. Cellular automata can achieve very complex emergent behaviors by simply using a few simple rules and communication with the immediate neighbours, therefore becoming an interesting candidate for the implementation of controllers.

\subsection{Approximate Computing}
Approximate and probabilistic computing have been recently proposed to increase energy-efficiency in fields where inexact results are tolerable \cite{Khasanvis2015a}. As discussed in \ref{appReq}, \acp{SDM} may fall into this category depending on the actual implementation of the metasurface pattern. This opens the door to a reduction of the voltage applied to the controller or the use of circuits providing approximate results in exchange for lower power. As long as the error probability remains bounded along the execution of the controller routines, this approach can reduce power consumption without noticeably degrading the performance of the \ac{SDM}. 

The metamaterial community can leverage existing knowledge in these areas, which have been applied across the computing stack: building approximate circuits, bounding the error probability throughout execution, debugging approximate devices, or combining the approach with energy harvesting, to name a few examples \cite{Han2013a, Venkatagiri2016, Xu2016, Mittal2016}.

\section{Applicability of On-chip Communication Techniques} 
\label{sec:contextNoC}
The system-level resemblance between multiprocessors and reconfigurable metamaterials suggest that on-chip communication techniques may be a valid approach for \ac{SDM}. As such, we next review a set of \ac{NoC} methods that could be applicable here. We make a distinction between wireline and wireless designs as it remains unclear which option is preferable \emph{a priori}: the wireless option avoids the use of conductive wiring which may interfere with the metamaterial plane, but comes at the expense of a higher complexity, i.e. the design and integration of tiny antennas and transceivers. 

\subsection{Network-on-Chip}
The \ac{NoC} paradigm essentially refers to packet-switched networks of integrated routers and links. In broad terms, research in this field has been mostly directed to scale designs while  obtaining high performance and reasonable efficiency. For high performance, objectives have been to minimize and bound latency in \acp{CMP} \cite{Grot2011, Krishna2014MICRO}, as well as to make better use of bandwidth in \acp{GPGPU} \cite{Bakhoda2010}. The main issue is that these proposals generally require fairly complex routers and wide links to implement their improvements and meet manycore requirements. Thus, they are not directly portable to the \ac{SDM} scenario. 

\acp{SDM} are much less sensitive to latency than \acp{CMP}, which automatically turns proposals seeking simplicity and low power into much better \ac{NoC}-based candidates for our target scenario. Next, we review several of these techniques. 

\noindento
{\bf Clockless NoC.} By default, most \ac{NoC} designs are clock-based. This requires the distribution of a clock signal throughout the chip, which takes precious area and power. To avoid it, one can adopt the \ac{GALS} approach consisting in the use of clockless links to communicate the cores \cite{Bjerregaard2005}. In a synchronous controller design, an interface is required to connect with the clockless network; whereas in an event-based approach, no further adaptation will be required. 

\noindento
{\bf Topology and router microarchitecture.} As in \acp{CMP}, a bidimensional mesh seems a natural fit for \acp{SDM} due to its ease of layout and performance. Yet still, even simpler topologies such as a ring \cite{Kim2009} are an intelligent choice since they allow the use of minimalistic router microarchitectures. In particular, the proposal by Kim {\em et al.} eliminates the need for both costly buffers to avoid losses and virtual channels to guarantee deadlock-freedom \cite{Kim2009a}. Another interesting point to consider here is whether clustering, i.e. serving groups of controllers via the same router, can help reduce footprint.

\noindento
{\bf Approximate communication.} The main idea behind approximate computing has been also applied to \acp{NoC}. Li \emph{et al.} proposed to use a lightweight lossy network to carry messages in program sections tolerant to errors \cite{Li2016}. Another approach would be to drop the supply voltage close to near-threshold levels, even if that results into occasional bit flips.  

\subsection{Wireless Network-on-Chip}
The \ac{WNoC} paradigm consists in the integration of antennas and transceiver circuits close to the computing cores, introducing higher flexibility at the network level \cite{Sujay2012, Laha2015}. Driven by the latency sensitivity and moderate throughput of \acp{CMP}, \acp{WNoC} are designed seeking high data rates and reasonable area. To this end, most proposals employ simple modulations such as \ac{OOK} and frequencies in the \ac{mmWave} range to obtain high bandwidth.

Again, the stringent constraints of \ac{SDM} suggest to sacrifice performance to reduce footprint. Since communication in \acp{SDM} is expected to be occasional and much less latency-sensitive than in \acp{NoC}, one can reduce the available bandwidth. This relaxes the requirements cast upon the antenna and transceiver and therefore enables the use of more compact circuits. Another technique that could be leveraged to reduce the footprint would be that of approximate computing: the main idea would be to reduce the gain of the power amplifier to save power even if that increases the bit error rate, as long as this error probability remains bounded within a safe margin. The use of electrically small antennas is another example of this footprint--performance tradeoff. 

\begin{figure}[!t]
\centering
\includegraphics[width=3.4in]{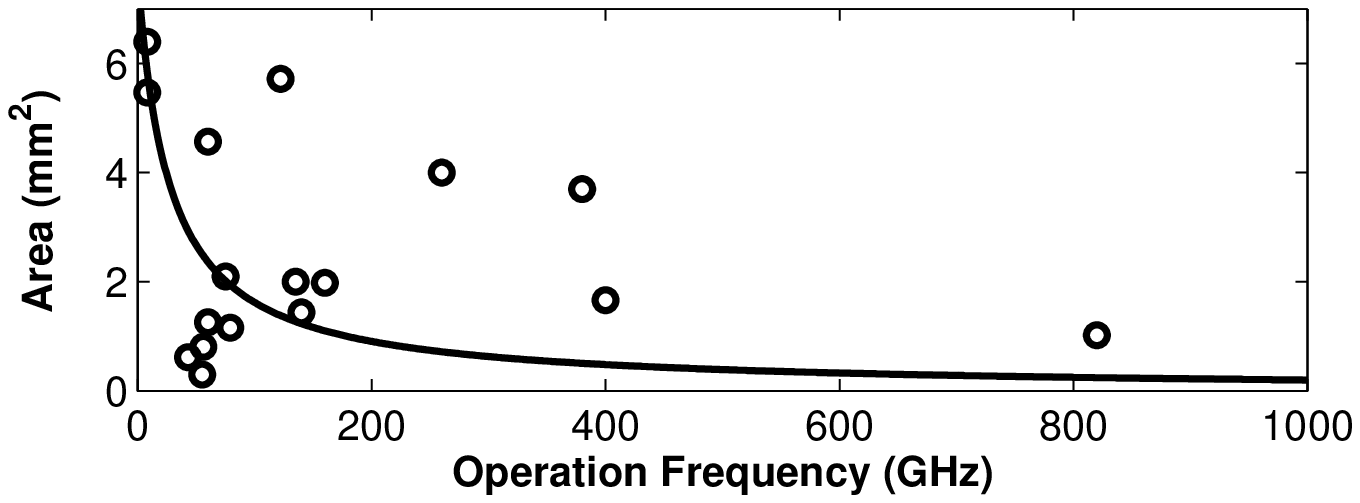}
\includegraphics[width=3.4in]{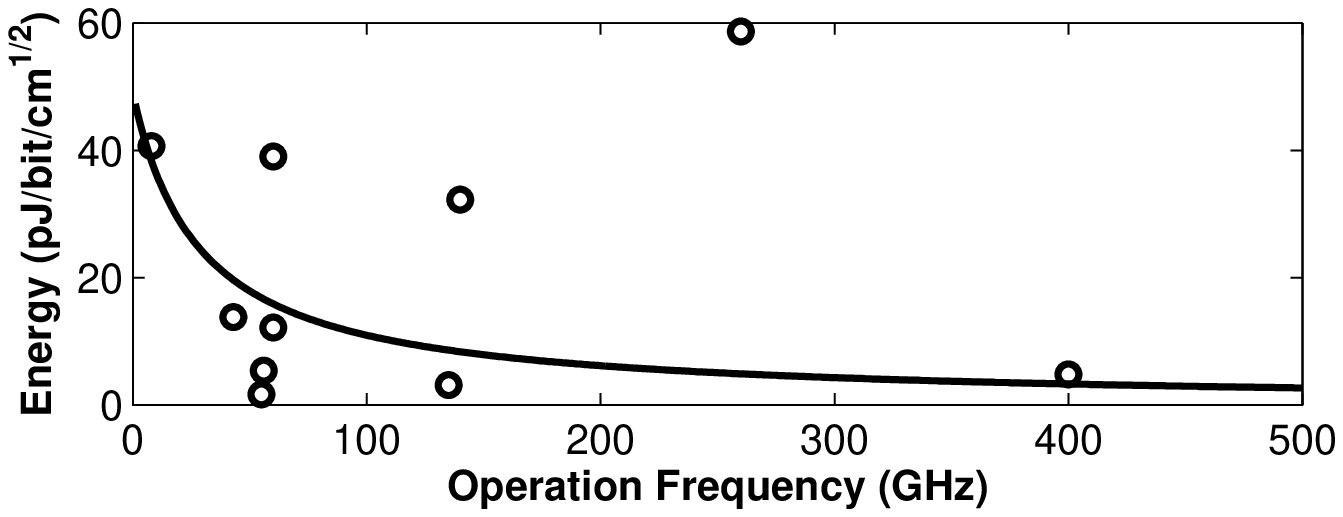}
\caption{Area and energy scalability of chip-scale wireless transceivers (data and fitting from \cite{AbadalTON}).}
\label{fig:area_energy}
\end{figure}

Although works assuming a large density of antennas within the same chip have been published \cite{Zhao2008, AbadalMICRO}, \acp{WNoC} generally complement a wireline \ac{NoC} and do not need many antennas to achieve meaningful results. The case for \ac{SDM}, however, is fundamentally different as the objective is to minimize wiring. This will probably require pushing the frequency used for communication up and beyond the \ac{mmWave} bands for two reasons: (1) to avoid coupling and interferences with the metamaterial plane, and (2) to achieve the target network density and efficiency, as both area and power scale inversely to frequency in on-chip environments (see Fig.~\ref{fig:area_energy}). 

The use of graphene-based antennas in the terahertz band can be a valid option for this particular purpose due to their outstanding properties \cite{AbadalTCOM}. The use of graphene as resonant sheets has been widely investigated, showing that patch or dipole antennas a few micrometers long and wide resonate in the terahertz band (0.1--10 THz), this is, between one and two orders of magnitude lower than their metallic counterparts \cite{Tamagnone2012, Jornet2013, Cabellos2014, Hosseininejad2016b}. Additionally, the unique tunability properties given by the relation between voltage bias and resonant frequency open the door to a set of new opportunistic communication protocols.

\section{Applicability of Nanonetworking Techniques}
\label{sec:contextOther}
As discussed in Section \ref{sec:infinitesimal}, the \ac{SDM} shows representative similarities with sensor and actuator networks. This suggests that ad hoc communication and networking mechanisms generally employed in such networks may be a candidate for the implementation of \acp{SDM}. In fact, the expected node density and huge physical constraints of intra-\ac{SDM} networks lead to considering extreme ad hoc solutions, which mostly lie in the nanonetworking domain \cite{Akyildiz2014}. 

Striving to maintain complexity at a minimum, most nanonetworking research finds consensus on the use of simplistic modulations such as the Time-Spread On-Off Keying (TS-OOK) \cite{Jornet2014OOK}. In TS-OOK, a logical 0 (1) is represented by means of a silence (short pulse), respectively, with a relatively long time between transmissions. This simplifies the receiver and reduces the probability of collisions. More over, this approach can be opportunistically combined with low weight coding \cite{Jornet2014CODING} and rate division multiple access \cite{Jornet2012a} to maximize its efficiency. 

Energy harvesting is another pillar of nanonetworking as it may enable the concept of perpetual networks. Its impact on the design of the protocol stack of nanonetworks has been under intense research over the last years, covering aspects such as the energy consumption policy \cite{Mohrehkesh2014} or the \ac{MAC} protocol \cite{Mohrehkesh2015} and assessing the potential network performance of perpetual networks \cite{Jornet2012}. The metamaterial community could benefit from these contributions since an important milestone is to make \acp{SDM} reconfigurable without compromising their autonomy. In particular, the work by Cid-Fuentes {\em et al.} \cite{Cid-fuentes2014}, which explores the design of energy harvesting systems in scenarios with high spatiotemporal traffic correlation, would be directly applicable to \acp{SDM} given the high expected correlation of traffic and potential harvesting sources in \acp{SDM}.

On top of all this, Liaskos {\em et al.} provided a view of the main networking challenges of \ac{SDM} and preliminary potential solutions from the nanonetworking point of view \cite{Liaskos2015}. The authors first discuss the problem of addressing in such dense networks and how it can be simplified taking into consideration the periodic, controlled, and monolithic nature of the system. As in \acp{NoC}, nodes can be unambiguously and statically identified with an internal id, leading to a major simplification of routing protocols \cite{Tsioliaridou2017} and a simplification or even complete elimination of addressing in particular case scenarios \cite{Tsioliaridou2016}. Finally, the authors propose the use of role-centric networking techniques, this is, defining custom roles in substitution of the conventional layered approach \cite{Liaskos2015a}. Preliminary evaluations were made for the data dissemination case (from interface to controllers), achieving a similar performance and an energy efficiency three times higher than with a generic protocol stack.

\section{Open Issues and Research Challenges}
\label{sec:challenges}
The \ac{SDM} design and optimization process poses new challenges for the various planes that comprise it. Envisioned milestones are detailed in the next subsections.

\subsection{Wireless channel characterization}
\label{subsec:Wireless-channel-characterizatio}
The communications plane constitutes the heart of the \ac{SDM}. The network of controllers is responsible for receiving external commands and finally altering the \ac{SDM} structure to meet a given objective. To this end, the efficiency of this network is critical: highly lossy communications may translate to redundant retransmissions of programmatic commands, resulting into higher \ac{SDM} setup times and reduced adaptivity potential. Thus, understanding and modeling the controllers' communication channel is critical for optimizing their communication accordingly.

The \ac{SDM} communications plane exhibits some unique attributes that affect the channel modeling. Specifically, the placement of the controllers is expected to exhibit a periodic layout, which is known to yield a well-defined chirality in the communication channel~\cite{Vegni.2016}. Additionally, the efficiency of the shielding plane is not a given, and may be subject to metamaterial plane restrictions. For instance, the presence of a highly conductive shielding layer underneath the metamaterial plane may result into a strong and unwanted reflection coefficient. Thus, a non-perfect shielding plane must be taken into account when studying the channel model, factoring for the interference from the metamaterial plane. This cross-talk can yield a highly non-linear channel, given that the programmatic commands exchanged by the controllers alter the metamaterial plane, in turn affecting the interference to the wireless channel. Note that most of these impairments are present in the physically similar \ac{WNoC} environment, for which comprehensive propagation models have not been developed yet \cite{Matolak2013CHANNEL, AbadalBOOK}.

\subsection{Abstracting the physics}
\label{subsec:Abstracting-the-physics}
\acp{SDM} are intended to be usable by non-physicists, which constitutes an attractive and challenging trait. In essence, an \ac{SDM} user should be able to define the required, high-level \ac{SDM} functionality without having to specify the low-level actions required to obtain it. Moreover, a user should be able to combine and multiplex \ac{SDM} functionalities, creating novel \ac{SDM} \emph{applications}. To these ends, the following \ac{SDM} software components need to be implemented: 
\begin{itemize} \itemsep1pt \parskip1pt \parsep1pt
\item An \ac{SDM} compiler, responsible for translating basic \ac{SDM} functionalities to the corresponding patterns that should be formed over its surface. These basic functionalities are those offered by metasurfaces in general, i.e., \ac{EM} absorption, steering, polarization, non-linear response~\cite{chen2016review}. The compiler essentially defines the low-level actions required to form these patterns, such as the state of switches in Fig.~\ref{fig:SDM}.  
\item An \ac{SDM} standard software library, offering the tools for monitoring, debugging, multiplexing and abstracting the basic \ac{SDM} functionalities towards higher-level objectives. For instance, an energy-harvesting high-level objective may be broken down to different \ac{EM} absorption commands per \ac{SDM} area unit. Monitoring software tools are required for establishing two-way communication with the \ac{SDM}, enabling for adaptive behavior and inter-connectivity within smart control loops. Finally, debugging tools are necessary for pinpointing both physical flaws (such as \ac{SDM} malfunction) and programming logic errors. 
\end{itemize}

From another point of view, these components constitute a software form of the physical laws governing the \ac{SDM} behavior. Three complimentary approaches are envisioned for accomplishing this transformation:
\begin{itemize} \itemsep1pt \parskip1pt \parsep1pt
\item \acp{SDM} can be treated as white-boxes, using existing analytical models of high-level objectives from the metamaterial world~\cite{PhysRevApplied.3.037001}. However, very few such models exist and their generality is limited. 
\item \acp{SDM} can be treated as black-boxes, and learning algorithms can be employed for correlating a high-level objective to a low-level \ac{SDM} internal state. Such algorithms examine multiple random \ac{SDM} configurations, converging to an understanding of their behavior. Nonetheless, this process can be computationally expensive and of limited efficiency.
\item \acp{SDM} can be treated as gray-boxes, empowering the learning algorithms with analytical insights to improve their efficiency.
\end{itemize}

Heuristics optimizers, such as genetic algorithms, may be used for yielding the optimal control plane state that best fits a sought \ac{EM} behavior \cite{Sean2011}. 

\subsection{Multi-physics simulation}
Optimizing the design of an \ac{SDM} via simulations pertains to its metamaterial and controller communication aspects. From the physics point of view, simulations are required for defining and optimizing the materials, dimensions, geometry and operating spectrum of the \ac{SDM}, and deducing the supported range of end-functionalities. From the communications point of view, the operational frequency and transmission power of nodes, their topology, allowed dimensions and materials need to be optimized, balancing minimal cross-talk with the metamaterial plane,
communication robustness and overall practicality. Additionally, joint physical/networking simulations are required for developing the \ac{SDM} software components outlined in Section~\ref{subsec:Abstracting-the-physics}.

Due to the aforementioned reasons, simulating \acp{SDM} is a necessary step in their design. However, it also constitutes a challenge on its own due to the dissimilarity of the two involved disciplines.  

The aspect of physics simulations commonly employs diverse computational and analytical methods (effective medium theories, FDTD, FEM, transfer matrix methods, heuristic algorithms, etc.) to study the \ac{EM} properties of the metamaterial plane. These techniques are known for their vast requirements in computational resources. The aspect of networking commonly operates at more abstract layers using discrete event simulators. Data packet-level propagation is considered sufficient for many networking systems, while statistical channel models simplify the simulation of the physical propagation medium.

Joining these two different aspects into one uniform simulator is an open challenge. Two possible resolutions are envisioned:
\begin{itemize} \itemsep1pt \parskip1pt \parsep1pt
\item Both aspects can be joined by a simulation at the physical layer. The periodicity in the \ac{SDM} geometry can be exploited for reducing the required computational resources to a tractable level. Specifically, node-pairs in identical or similar surroundings can be simulated once and then be cached and re-used for the duration of the simulation.
\item The two aspects are kept separate, with the more abstract networking events driving the low-level physical layer simulation. The network communication channel is treated statistically, as described in Section~\ref{subsec:Wireless-channel-characterizatio}.
\end{itemize}

In both cases, it is noted that latest computational methods taking advantage of multiple CPUs and GPUs have exhibited several orders of magnitude shorter simulation times for the physics aspect of this challenge~\cite{kantartzis2016generalized}.

\section{Conclusions}
\label{sec:conclusions}
\acp{SDM} are expected to overcome the main limitations of conventional metamaterials in terms of reusability, adaptivity, and accessibility to the engineering community. The materialization of this vision requires embedding a network of tiny controllers within the metamaterial structure, which represents an important challenge due to the particularities of the application context. On the one hand, we have identified the planar, integrated, and monolithic nature of \ac{SDM} as characteristics suggesting to treat this application as a scaled version of a manycore embedded system with a NoC, either wired or wireless. On the other hand, its constrained and ultra-dense landscape, as well as the event-based and correlated nature of the workload, brings \acp{SDM} closer to the nanosensor network scenario. A graceful combination of both top-down and bottom-up design approaches may lead to a unique, custom solution meeting the demands of this new disruptive paradigm.

\section*{Acknowledgment}
This work was funded by the European Union via the Horizon 2020: Future Emerging Topics call (FET Open), grant EU736876, project VISORSURF (http://www.visorsurf.eu).

\ifCLASSOPTIONcaptionsoff
  \newpage
\fi

%

\begin{IEEEbiography}[{\includegraphics[width=1in,height=1.25in,clip,keepaspectratio]{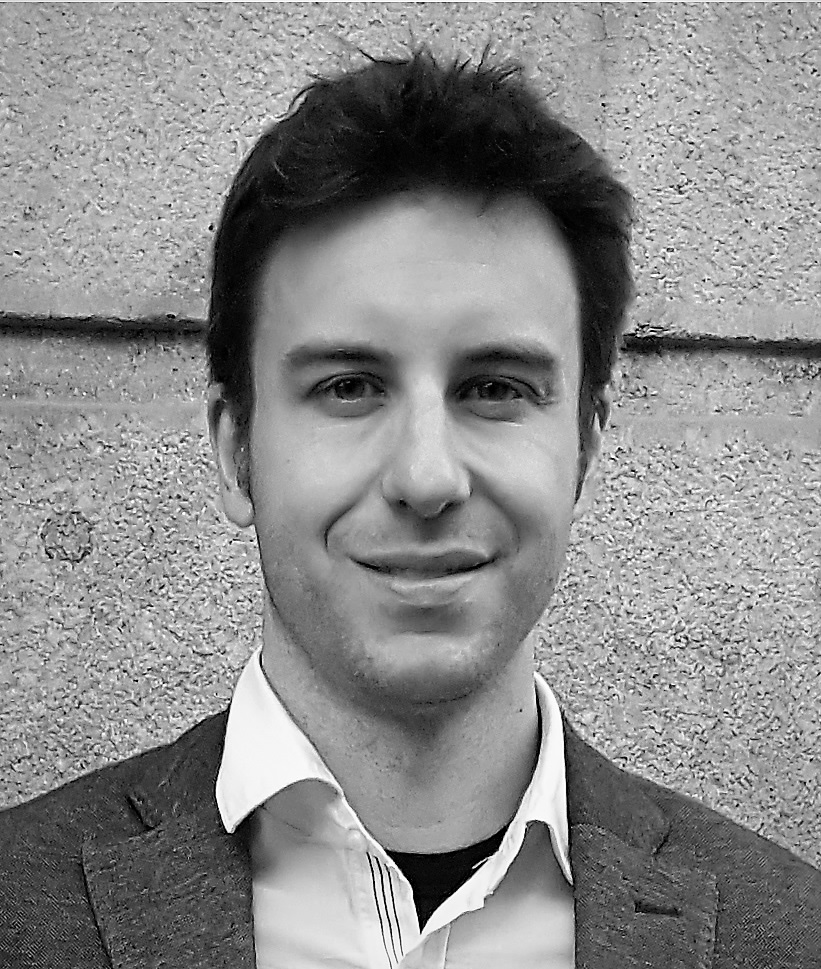}}]{Sergi Abadal}
received the BSc and MSc degrees in telecommunication engineering from the Universitat Polit\`{e}cnica de Catalunya (UPC), Barcelona, Spain, in 2010 and 2011, and the PhD in computer architecture from the same institution in 2016. Since then, he works as a postdoctoral researcher at the NaNonetworking Center in Catalunya (N3Cat). From 2009 to 2010, he was a Visiting Researcher with the Broadband Wireless Networking Laboratory, Georgia Institute of Technology, Atlanta, USA. From May to November 2015, he was a visiting researcher at the i-acoma group, School of Computer Science, University of Illinois, Urbana-Champaign. His current research interests are ultra-high-speed on-chip wireless networks and broadcast-enabled manycore processor architectures. Dr. Abadal was awarded by INTEL within its Doctoral Student Honor Program in 2013.
\end{IEEEbiography}

\begin{IEEEbiography}[{\includegraphics[width=1in,height=1.25in,clip,keepaspectratio]{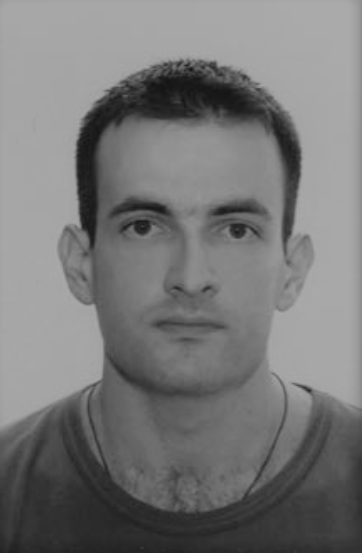}}]{Christos Liaskos}
received the Diploma in Electrical and Computer Engineering from the Aristotle University of Thessaloniki (AUTH), Greece in 2004, the MSc degree in Medical Informatics in 2008 from the Medical School, AUTH and the PhD. degree in Computer Networking from the Dept. of Informatics, AUTH in 2014. He has published work in several venues, such as IEEE Transactions on: Networking, Computers, Vehicular Technology, Broadcasting, Systems Man \& Cybernetics, Networks and Service Management, Communications, INFOCOM. He is currently a researcher at the Foundation of Research and Technology, Hellas (FORTH). His research interests include computer networks and nanotechnology, with a focus on developing nanonetwork architectures and communication protocols for future applications.
\end{IEEEbiography}

\begin{IEEEbiography}[{\includegraphics[width=1in,height=1.25in,clip,keepaspectratio]{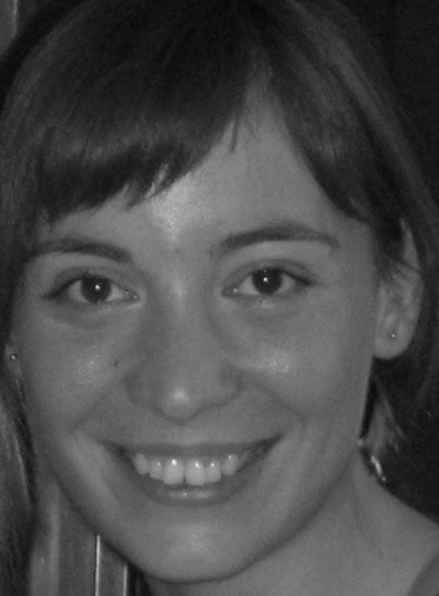}}]{Ageliki Tsioliaridou}
received the Diploma and PhD degrees in Electrical and Computer Engineering from the Democritus University of Thrace (DUTH), Greece, in 2004 and 2010, respectively. Her research work is mainly in the field of computer networks and specific focus on the problem of Quality of Service. Additionally, her recent research interests lie in the area of nanonetworks, with specific focus on architecture, protocols and security/authorization issues. She has contributed to a number of EU, ESA and National research projects. She is currently a researcher at the Foundation of Research and Technology, Hellas (FORTH).
\end{IEEEbiography}

\begin{IEEEbiography}[{\includegraphics[width=1in,height=1.25in,clip,keepaspectratio]{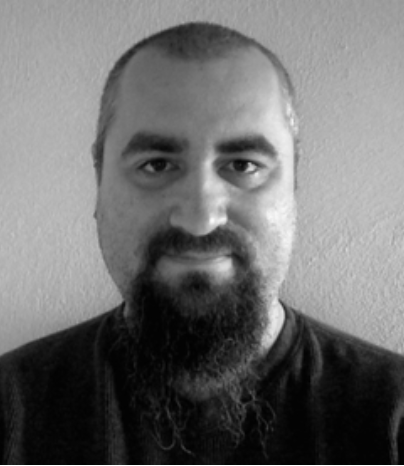}}]{Sotiris Ioannidis}
(male) received a BSc degree in Mathematics and an MSc degree in Computer Science from the University of Crete in 1994 and 1996 respectively. In 1998 he received an MSc degree in Computer Science from the University of Rochester and in 2005 he received his PhD from the University of Pennsylvania. Ioannidis held a Research Scholar position at the Stevens Institute of Technology until 2007, and since then he is Research Director at the Institute of Computer Science of the Foundation for Research and Technology - Hellas. His research interests are in the area of systems, networks, and security. Ioannidis has authored more than 100 publications in international conferences and journals, as well as book chapters, including ACM CCS, ACM/IEEE ToN, USENIX ATC, NDSS, and has both chaired and served in numerous program committees in prestigious international conferences. Ioannidis is a Marie-Curie Fellow and has participated in numerous international and European projects. He has coordinated a number of European and National projects (e.g. PASS, EU-INCOOP, GANDALF) and is currently the project coordinator of the SHARCS and CYBERSURE H2020 European projects.
\end{IEEEbiography}

\begin{IEEEbiography}[{\includegraphics[width=1in,height=1.25in,clip,keepaspectratio]{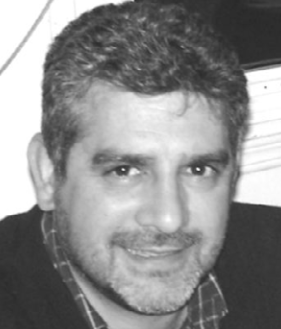}}]{Andreas Pitsillides}
is a Professor in the Department of Computer Science, University of Cyprus, and heads the Networks Research Laboratory he founded in 2002. Since May 2014 he is a Visiting Professor at the University of Johannesburg. His research interests include communication networks, the Internet- and Web- of Things, Smart Spaces (Home, Grid, City), and Nanonetworking. He has published over 250 referred papers in flagship journals (e.g. IEEE, Elsevier, IFAC, Springer), international conferences, and books. He serves on the editorial board of the Journal of Computer Networks (COMNET) and served on several international conferences as general chair, international co-chair, technical program chair, and on executive committees, technical committees, guest co-editor and invited speaker. He has participated in over 35 EU and locally funded research projects.
\end{IEEEbiography}

\begin{IEEEbiography}[{\includegraphics[width=1in,height=1.25in,clip,keepaspectratio]{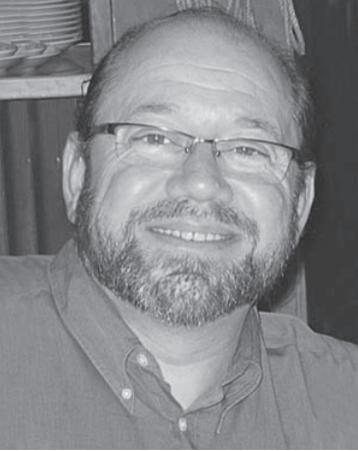}}]{Josep Sol\'{e}-Pareta}
obtained his MSc degree in Telecom Engineering in 1984, and his PhD in Computer Science in 1991, both from the Universitat Polit\`{e}cnica de Catalunya (UPC). In 1984 he joined the Computer Architecture Department of UPC. Currently he is Full Professor within this department. He did a Postdoc stage (summers of 1993 and 1994) at the Georgia Institute of Technology. He is co-founder of the UPC-CCABA (http://www.ccaba.upc.edu), and UPC-N3cat (http://www.n3cat.upc.edu). His current research interests are in Nanonetworking Communications, Traffic Monitoring and Analysis, High Speed and Optical Networking, and Energy Efficient Transport Networks, with emphasis on traffic engineering, traffic characterization, MAC protocols and QoS provisioning. His publications include several book chapters and more than 300 papers in relevant research journals ($>$80), and refereed international conferences. He has participated in many European projects in the Computer Networking field. He was Local Chairman of the 25th Conference on Computer Communications (IEEE INFOCOM 2006) held in Barcelona on April 23-29, 2006, and General Chairman of the 7th International Conference on Transparent Optical Networks (ICTON 2005, Barcelona, July 3-7, 2005), and of the 5th Workshop on Quality of Future Internet Services (QofIS’04, Barcelona, September 29-30 and October 1, 2004). 
\end{IEEEbiography}

\begin{IEEEbiography}[{\includegraphics[width=1in,height=1.25in,clip,keepaspectratio]{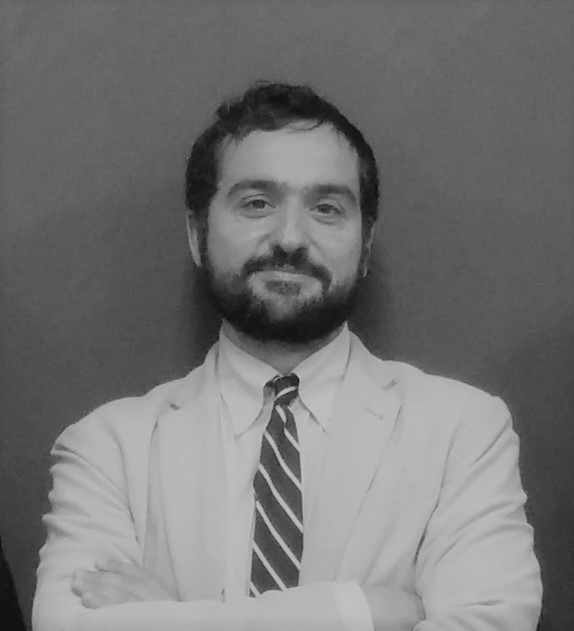}}]{Eduard Alarc\'{o}n}
received MSc (national award) and PhD degrees in EE from Universitat Polit\`{e}cnica de Catalunya (UPC), Spain, in 1995 and 2000, respectively, where he became Associate Professor in 2001, and has been visiting Professor at University of Colorado at Boulder, USA (2003, 2006, 2008) and KTH Stockholm (2011). He has coauthored more than 300 scientific publications, 8 book chapters and 8 patents, and has been involved in different national, EU and US R\&D projects. Research interests include the areas of on-chip energy management circuits, energy harvesting and wireless energy transfer, nanocommunications and small satellites. He has been funded and awarded several research projects by companies including Google, Samsung and Intel. He has given 30 invited lectures and tutorials worldwide. He is Vice President of the IEEE CAS society, was elected member of the IEEE CAS Board of Governors (2010-2013) and was IEEE CAS society distinguished lecturer, recipient of the Best paper award at IEEE MWSCAS98, co-editor of 6 journal special issues, 8 conference special sessions, TPC co-chair and TPC member of 30 IEEE conferences, and Associate Editor for IEEE TCAS-I, TCAS-II, JETCAS, JOLPE and Nano Communication Networks. 
\end{IEEEbiography}

\begin{IEEEbiography}[{\includegraphics[width=1in,height=1.25in,clip,keepaspectratio]{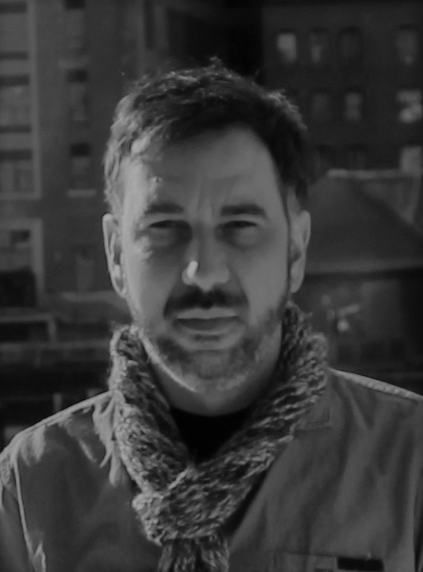}}]{Albert Cabellos-Aparicio}
received the BSc, MSc, and PhD degrees in computer science engineering from the Universitat Polit\`{e}cnica de Catalunya (UPC), Barcelona, Spain, in 2001, 2005, and 2008, respectively. He has also been an Assistant Professor with the Computer Architecture Department and Researcher with the Broadband Communications Group, Technical University of Catalunya, since 2005. In 2010, he joined the NaNoNetworking Center in Catalunya, where he is the Scientific Director. He is an Editor of Nano Communication Networks and founder of the ACM NANOCOM conference, the IEEE MONACOM workshop, and the N3Summit. He has also founded the LISPmob open-source initiative along with Cisco. He has been a Visiting Researcher with Cisco Systems, San Jose, CA, USA, and Agilent Technologies, Santa Clara, CA, USA, and a Visiting Professor with the Royal Institute of Technology (KTH), Stockholm, Sweden, and the Massachusetts Institute of Technology (MIT), Cambridge, MA, USA. He has given more than 10 invited talks (MIT, Cisco, INTEL, MIET, Northeastern University, etc.) and coauthored more than 15 journal and 40 conference papers. His main research interests are future architectures for the Internet and nanoscale communications.
\end{IEEEbiography}



%
%
%
%
%
%

\end{document}